\documentstyle[12pt]{article}

\font\sqi=cmssq8
\def\DR{\rm I\kern-1.45pt\rm R}
\def\DC{\kern2pt {\hbox{\sqi I}}\kern-4.2pt\rm C}
\renewcommand{\thefootnote}{\fnsymbol{footnote}}
\begin{document}
\thispagestyle{empty}
\begin{center}
{\Large\bf A note on quantum Bohlin transformation}
 \vspace{1.5cm}
 \\
{\large A. Nersessian}\footnote{e-mail:nerses@thsun1.jinr.dubna.su},
{\large V. Ter-Antonyan}\footnote{e-mail:terant@thsun1.jinr.dubna.su},
 {\large M.Tsulaia}\footnote{e-mail:tsulaia@thsun1.jinr.dubna.su}
\vspace{0.5cm}\\
{\it Bogolyubov Laboratory of Theoretical Physics,} \\
{\it Joint Institute for Nuclear Research}\\
{\it Dubna, Moscow region, 141980, Russia}
 \end{center}
\vspace*{1cm}
\begin{abstract}
It is shown, that  the  reduction  of
the circular quantum oscillator  by the  $Z_2$ group action
results  to the  two systems:
a two-dimensional hydrogen atom,  and a  ``charge - charged magnetic
vortex"  one, with the  spin $\frac 12$.
 Analogously,  the $Z_N$ reduction of the  two-dimensional
system with the central potential $r^{2(N-1)}$
results into $N$  bound  ``charge -
 magnetic vertex" systems with
the interaction potential $r^{2(1/N-1)}$ and
spins $\sigma=\frac kN$, $k =0,1,..., (N-1)$.  \end{abstract}
\centerline{{\it PACS number : 03.65.Ge }}
\setcounter{page}0
\renewcommand{\thefootnote}{\arabic{footnote}}
 \setcounter{footnote}0
\newpage
 \setcounter{equation}0
{\bf 1. Introduction.}
As it was established by Bohlin \cite{Bohlin}, trajectories of the
circular oscillator and the two-dimensional Kepler problem
are connected by the transformation
 \begin{equation}
w=z^2 ,
\label{bo}\end{equation}
where the complex coordinates $z$ and $w$ parametrize the position of a
particle in the oscillator and Kepler problem.

Due to  (\ref{bo}), to the  tracing along the ellipse in the oscillator
problem there corresponds the double tracing of the ellipse in the Kepler
problem.
 In the quantum case, this is reflected in the fact,
that the circular oscillator is transferred by (\ref{bo})
 into an (non-physical)
Coulomb problem on the  {\it two-sheet Riemann surface}.

In this note we show that if prior to the transformation (\ref{bo})
 we perform the reduction of the oscillator by the $Z_2$ group action,
 given the parity operator, the oscillator splits into two physical systems:
a two-dimensional hydrogen atom (even states) and a bound
``charge--charged magnetic vortex" system (odd states).
The second system has the eigen angular momentum (spin) $\frac 12$
and its energy levels are shifted with respect to
the levels of the first system.

We have also shown that the two-dimensional system with the potential
$|z|^{2(N-1)}$ after the reduction by the $Z_N$ group action
 and subsequent  transformation $w=z^N$ splits into
$N$ "charge-- magnetic vortex" systems, interacting
with the potential  $|w|^{-2(1-1/N)}$, which have eigen  angular
momenta $\sigma=\frac kN$, $k =0,1,..., (N-1)$ .\\

{\bf 2.Bohlin transformation.}
 Let us consider the Schr\"odinger equation for
the isotropic oscillator on the plane $\dot{\DC}=\DC -\{0\}$
\begin{equation}
4\partial_z{\bar\partial}_{ z}\Psi+\frac{2\mu}{\hbar^2}
(E-\frac{\mu\omega^2}{2} z\bar z)\Psi=0,
\label{1}\end{equation}
where $\partial_{z}={\partial}/{\partial{z}}$,
${\bar\partial}_{z} ={\partial}/{\partial{\bar z}}$.

Reduce Eq.(\ref{1}) by the group $Z_2$,
  having restricted ourselves to even  ($\sigma=0$)
or odd ($\sigma=\frac 12$)  solutions  of Eq. (\ref{1})
 \begin{equation}
  \Psi_\sigma (z, \bar z) =
\psi_\sigma (z^2, {\bar z}^2){\rm e}^{2i\sigma {\rm arg}\;z},
\label{3}\end{equation}
and then perform the transformation
 \begin{equation}
w=\lambda z^2 , \quad \lambda=(\frac{\mu\omega}{\hbar})^\frac 12 .
\label{b}\end{equation}
Let us call a set of these transformations {\it the quantum Bohlin
transformation} (note that in contrast with formula (\ref{bo})
 the coordinates  $z$ and $w$ in formula (\ref{b}) have the same dimension).

According to Eq.(\ref{3}), the wave functions  $\psi_\sigma$
satisfy the condition $$\psi_{\sigma}(|w|, {\rm arg} w + 2\pi)=
\psi_{\sigma}(|w|, {\rm arg} w ),$$ which imply that the range  of
definition ${\rm arg}\;w\in[0,4\pi)$ can be restricted,
without loss of generality, to ${\rm arg} w\in[0,2\pi)$.

Thus, the quantum Bohlin transformation  reduces the oscillator
to the system on the plane rather than on
the two-sheet Riemann surface.

One can easily see that the quantum Bohlin
transformation reduces eq.(\ref{1})
to the form
\begin{equation}
\frac{1}{2\mu} {\hat p}_\sigma {\hat p}^{+}_\sigma\psi(w,\bar w)
-(\frac{\alpha}{|w|} +{\cal E} )\psi_\sigma=0, \label{4}
\end{equation}
where
$${\hat p}_\sigma = -2i\hbar\partial_w-\frac{i\hbar\sigma}{2w},$$
and the parameters ${\cal E} $ and $\alpha$
are determined by the
conditions
\begin{equation} {\cal E}=-\frac{\mu\omega^2}{8\lambda^2}, \quad
\alpha=\frac{E}{4\lambda}.
\label{parameter}\end{equation}
and have the meaning of the energy and coupling constant of the system
obtained.

Equation (\ref{4}) can be interpreted as the Schr\"odinger equation of a
nonrelativistic particle with electric charge  $e$ in the static
electromagnetic field, determined by  the potential
\begin{equation}
{\cal A}=(\phi, A_w):\quad \phi=
-\frac{Q}{|w|},\quad A_w=\frac{ig}{w},
\label{Aphi}\end{equation}
 where
\begin{equation} ¥Q={\alpha},\quad\frac{eg}{\hbar c}=\sigma =0,\frac{1}{2}.
\label{qd}\end{equation}
It is to be noted that  $\phi$ is the potential
of the two-dimensional Coulomb field of a particle with charge $Q$, and
 $A_w$ is a vector potential of an
infinitely thin solenoid-- flux tube ( magnetic vortex, in two
dimensional interpretation): to it there corresponds zeroth
strength of the magnetic field $B=rot A_w=0$ ($w\in\dot{\DC}$)
 and nonzero magnetic flux $2\pi g$.

Consequently, the potential ${\cal A}=(\phi, A_w)$
defines the electromagnetic field the
two-dimensional charged magnetic vortex (two-dimensional dyon) with the
electric $Q$ and magnetic  $g$  charges.

Thus, having reduced the two-dimensional oscillator by the $Z_2$ group action,
 we have obtained a two-dimensional hydrogen atom at $\sigma=0$  and a
"charge-charged magnetic vortex" system ("charge--two-dimensional dyon") at
$\sigma =\frac 12$ . \\

The Hamiltonian of the circular oscillator has hidden symmetry $so(3)$
\begin{equation}
\{ I,I^{+} \}=-2(\hbar\omega)^2 J,
\quad  \{ I, J \}=-2I,\quad  \{ I^+, J\}=2I^+
     \label{so3}\end{equation}
generated by the angular momentum operator
     \begin{equation} J= 2(z\partial-\bar z\bar\partial )
     \label{j0} \end{equation}
and the vector (complex)   constant of motion
 \begin{equation}
     I=-\frac{2\hbar^2}{\mu}{\bar\partial}_z^2+
\frac{\mu\omega^2}{2}{ z}^2.
     \label{rl0} \end{equation}
These operators are even ($Z_2$-invariant) and,
therefore, the reduced systems
also have hidden  $so(3)$ symmetry that is generated
by these operators
reduced by the quantum Bohlin transformation to the form
\begin{eqnarray} &J\to &2J_\sigma
,\;\;J_\sigma=\frac{i}{\hbar}\left( w{\hat p}_{\sigma}- {\bar w}{\hat
p}^+_{\sigma}\right) \label{j} \\
& I\to & 4\lambda{I}_\sigma\;\;{I}_\sigma=
\frac{i}{2\mu}(J_\sigma{\hat p}_\sigma+ {\hat p}_\sigma J_\sigma)
 -\frac{\alpha w}{|w|}.
 \label{rl}\end{eqnarray}
  Note that ${J}_\sigma$  is the angular
momentum operator of the reduced system and ${I}_\sigma$ is its Runge--Lenz
 vector.

It is seen from expression  (\ref{j})
that the eigenvalues  of the angular momenta
  of the reduced system  and the oscillator,
$m_\sigma$  and $M$, are related   by the expression $M= 2m_\sigma$,
from which   follows
 \begin{equation} m_\sigma= \pm\sigma, \pm (1+\sigma),
\pm (2 +\sigma),\ldots .
 \end{equation}
  Using expressions (\ref{parameter}), we  easily get from
 the oscillator energy spectrum
 \begin{equation}
E=\hbar\omega(2N_r+ |M|+1), N_r=0, 1, 2,...,
  \label{eo}\end{equation}
  the energy spectrum  of the reduced system
 \begin{equation} {\cal E}_\sigma=
 -\frac{\mu\alpha^2}{2\hbar^2 (N_r+|m_\sigma|+\frac 12)^2}.
  \end{equation}
Hence, the ``charge-charged magnetic vortex" system has the eigen angular
 moment $ \sigma= \frac 12$ and the energy spectrum is shifted
 with respect to  the energy spectrum of the
 two- dimensional hydrogen atom , i. e. the Aharonov - Bohm effect
takes place.

Note also that composites ``charge - magnetic vortex" are
  {\it anyons}, particles with fractional statistics and spin
equal to their eigen angular momentum  (see, e.g.  \cite{wilchek}).\\

Finally, using expression (\ref{3}), we find from the wave functions
 of the oscillator
   \begin{equation}
 \Psi_{N_r, M}(z,\bar z)= C_{N_r,|M|} (\lambda|z|)^{|M|}
{\rm e}^{-\frac{\lambda^2|z|^2}{2}}L_{N_r}^{|M|}(\lambda^2|z|^2)
{\rm e}^{iM {\rm arg} z}/{\sqrt 2\pi} .
 \label{wo}\end{equation}
  the wave functions of the reduced system
\begin{equation}
\psi_{\sigma, N_r, m_\sigma}=
\frac {C_{N_r, 2|m_\sigma|}}{2{\sqrt{N_r+|m_\sigma|+ \frac 12}}}
(\lambda_{N_r,m_\sigma} |w|)^{|m_\sigma|}
{\rm e}^{-\frac{\lambda_{N_r, m_\sigma}|w|}{2}}
L_{N_r}^{2|m_\sigma|}(\lambda_{N_r, m_\sigma} |w|)
{\rm e}^{i(m_\sigma -\sigma){\rm arg} w}/{\sqrt 2\pi} .
\end{equation}
Here
 $L_{p}^s(x)$  are the generalized Laguerre polynomials, related with the
confluence hypergeometric function by
$$L_{p}^s(x)= \frac{\Gamma(s+p+1)}{\Gamma(s+1)} F(-p, s+1, x),$$
and the constants $C_{N_r,|M|}$  and $\lambda_{N_r, m_\sigma}$
have the form
$$C_{N_r,|M|}=\frac{{\sqrt 2}\lambda}{{\sqrt{ N_r!(N_r+|M|)!}}},
\quad \lambda_{N_r, m_\sigma}=
\frac{2}{r_B (N_r+m_\sigma+ \frac 12)}.$$
 where $r_B =\frac{\hbar^2}{\mu\alpha}$ is the ``Bohr radius" of a hydrogen
atom.\\

{\bf 3. Generalization.} In a popular book by Arnold \cite{arnold}
the generalization of the Bohlin transformation
\begin{equation}
w=z^N,
\label{ab}\end{equation}
is given,  connecting the classical trajectories   of the two-dimensional systems
with the potentials  $|z|^{2a}$  and  $|w|^{2b}$ , where
 \begin{equation}
(a+ 1)(b + 1)=1 ,\quad N= a+1.
 \label{abn}\end{equation}

What does the transformation (\ref{ab}) lead to in the quantum case?\\
We restrict ourselves, for simplicity, to the case, when the parameter  $N$
is a natural number.

Let us consider on the complex plane $\dot{\DC}$
the spectral problem (here and hereafter  $\hbar=\mu=c=1$)
\begin{equation}
\left\{
\begin{array}{rcl}
&(-2\partial_z{\bar\partial}_z+
A|z|^{2(N-1)} -E)\Psi=0&\\
&2(z\partial_z- {\bar z}{\bar\partial_z})\Psi=M\Psi& ,
\end{array}\right.
\label{111}\end{equation}
with the single-valued  wave functions
$$\Psi(z,{\bar z})=R_{M, E}(|z|){\rm e}^{iM {\rm arg} z}\quad
M=0,\pm 1, \pm 2,\ldots . $$
The transformation (\ref{ab}) maps the system \ref{111}
 into the one  on the $N$-sheet  Riemann surface.
 Therefore, reduce the
system (\ref{111}) by the group $Z_N$, substituting
 the wave functions of the form
\begin{equation} \Psi_\sigma (z, \bar z) = \psi_\sigma
 (z^N, {\bar z}^N) {\rm e}^{Ni\sigma {\rm arg}\;z}, \quad \sigma=\frac kN,
\quad k=0,1,...(N-1).
 \label{20} \end{equation}
 into the system of equations (\ref{111}) and  then performed the
transformation  (\ref{ab}),  we get
\begin{equation} \left\{ \begin{array}{rcl}
&(\frac 12 {p_\sigma {\bar p}_\sigma}
+\alpha|w|^{2(\frac 1N -1)} - {\cal E})\psi(w,\bar w)=0&\\
&2i(wp_\sigma-{\bar w}{\bar p}_\sigma)\psi=m_\sigma\psi &
\end{array}\right.
\nonumber\end{equation}
where
\begin{equation}
p_\sigma= -2i\partial_w -\frac{i\sigma}{w},\quad {\cal E}=-\frac{A}{N^2},
\quad \alpha= -\frac{E}{N^2}  ,
\end{equation}
and the eigenvalues of the angular momentum of the system     are
$$m_\sigma = \pm \sigma , \pm(1+ \sigma), \pm (2+ \sigma),\ldots .$$
Thus, as a result of the reduction by the group $Z_N$
the system (\ref{111})  disintegrated into $N$ bound
``charge-magnetic vortex" systems with the interaction potential
$\beta |w|^{-2(\frac 1N-1)}$ and eigen angular momenta
 $\sigma=0, 1/N, 2/N,\ldots,(N-1)/N$.  \\

{\bf 4. Conclusion}
 We have shown that the quantum Bohlin transformation splits the
two-dimensional oscillator into two systems having the Coulomb symmetry:
the two-dimensional hydrogen atom and the ``charge-charged magnetic vortex"
system with a half-integer angular momentum. However, this is not the only
connection between the oscillator and topologically
 nontrivial quantum systems with hidden Coulomb symmetry.
Thus, the reduction of the four-dimensional oscillator by the group $U(1)$
together with the Kuustanheimo-Steifel \cite{tn}
transformation results in the three-dimensional
bound system ``charge-$U(1)$-dyon" \cite{Z} that is described by the
Hamiltonian
$$
{\hat H}=\frac{\hbar^2}{2\mu} {\hat\pi}^2  +\frac{\hbar^2s^2}{2\mu r^2} -
\frac{\alpha}{r},\quad s= 0,\pm 1/2, \pm 1, \ldots,
$$
where
$$
[{\hat\pi}_i, {\hat\pi}_j]= is\epsilon_{ijk}\frac{x_k}{r^3} , \quad
 [{\hat\pi}_i, x_j]=-i\delta_{ij}  $$
and ${\bf r}/{r^3}$ is the magnetic field strength of the   $U(1)$-monopole
with the unit magnetic charge.
      Note that the Hamiltonian  of the system
  (\ref{4})  can be represented in an analogous form
$${\hat H}_\sigma=-\frac{4\hbar^2}{2\mu}{\partial_w}{\bar\partial}_w
+\frac{\hbar^2\sigma^2}{2\mu|w|^2} -\frac{\alpha}{|w|}.$$

{\bf Acknowledgement.} We are grateful to V. Osipov for useful discussions.

\end{document}